# Functional differentiation under simultaneous conservation constraints*

Tamás Gál

Department of Theoretical Physics, University of Debrecen, H-4010 Debrecen, Hungary
E-mail: galt@phys.unideb.hu

**Abstract:** Analytical formulae for functional differentiation under simultaneous $K$-conservation constraints, with $K$ the integral of some function of the functional variable, are derived, making the proper account for the simultaneous conservation of normalization and statistical averages, e.g., possible in functional differentiation in nonvariationally built physical theories, which gets particular relevance for nonequilibrium, time-dependent theories.

Functional differentiation appears as a basic constituent of physical theories ranging from hydrodynamics to quantum field theories [1]. In many of the cases, constraint ($C[\rho] = 0$), for example the conservation law of some extensive property, limits the possible changes described by the given physical theory, which leads to a potential shift of the functional derivative (that governs the changes to first order) from its unconstrained form $\frac{\delta A[\rho]}{\delta \rho(x)}$ in the physical equation. The shift appears in the form of a "Lagrange multiplier" $\mu$ (a multiplier constant in *x*, which becomes the Lagrange multiplier at the extrema of $A[\rho]$ with the constraint),

$$\frac{\delta A[\rho]}{\delta \rho(x)} - \mu \frac{\delta C[\rho]}{\delta \rho(x)} , \qquad (1)$$

with $\mu$ only partly (and implicitly) determined by the physical equation and the constraint, apart from the case of an Euler-Lagrange equation. A general (that is, independent of the physical problem the given constrained derivative appears in), explicit analytical determination of multipliers $\mu$ becomes important with respect to (primarily, time-dependent, nonequilibrium) physical theories where the physical equations do not emerge, at least not directly, variationally, from some variational theorem, as functional derivatives. However, the method of that treatment of constraints had not been known until recently, as a consequence of which the account for constraints had been possible only in a *direct* variational way even in nonstationary problems [2], limiting the construction of practical physical theories with constraints. In [3,4], for the important class of constraints

$$\int f(\rho(x))dx = K \qquad (2)$$

(where *f* is an invertible function, and an explicit *x*-dependence of *f* is allowed as well, though not denoted for simplicity), an analytical formula for constrained derivatives, Eq.(1), has been derived:

$$\frac{\delta A[\rho]}{\delta_K \rho(x)} = \frac{\delta A[\rho]}{\delta \rho(x)} - \frac{f^{(1)}(\rho(x))}{K} \int \frac{f(\rho(x'))}{f^{(1)}(\rho(x'))} \frac{\delta A[\rho]}{\delta \rho(x')} dx' , \tag{3}$$

giving a treatment of functional differentiation under conservation constraints including the simple normalization conservation $\int \rho(x)\,dx = N$, the conservation of statistical averages

$$\int g(x)\rho(x)dx = L \tag{4}$$

(that is, linear constraints $L[\rho] = L$, with $L[\delta(x'-x)] = g(x)$), or entropy conservation, $\int -k\rho(x)\ln\rho(x)dx = S$, e.g. Very recently, this constrained differentiation has been applied by Clarke [5] to build a dynamical model of simultaneous dewetting and phase separation in thin-film binary mixtures [6], through the proper account for conservation constraints. The use of Eq.(3) and of the method behind it, however, is limited, being applicable only for one constraint on a single functional variable. In the physical areas where an explicit handling of constrained functional derivatives can be of particular relevance, there may be more than one simultaneous conservation requirement constraining the variation of functional variables, as in statistical physics [7] or in the physics of complex systems (e.g., in liquid film dynamics [8]); so the extension of the idea of the formula Eq.(3) is essential for its general physical use. In this paper, the generalization of Eq.(3) for functional differentiation under multiple $K$-conservation constraints, embracing simultaneous conservation of normalization and some statistical average, e.g., will be set up.

The extension of Eq.(3) is not trivial, as the successive application of two $K$-conserving projection operators $\hat{p}_{K_1}$ and $\hat{p}_{K_2}$, defined by $\hat{p}_K(x,x')h(x') = \int dx' \frac{\delta \rho(x)}{\delta_K \rho(x')} h(x')$, does not yield a $(K_1, K_2)$-conserving projection, that is, $\hat{p}_{K_1,K_2} \neq \hat{p}_{K_1}\hat{p}_{K_2}$, meaning that $\hat{p}_{K_1}\hat{p}_{K_2}\delta\rho$ is not a $(K_1, K_2)$-conserving first-order variation. ($\hat{p}_{K_1}$ and $\hat{p}_{K_2}$ do not even commute.) To obtain a formula for $(K_1, K_2)$-conserving (or

-constrained) differentiation, the definition of *K*-constrained derivatives described in Sec.4 of [9] will be taken as basis. In [9], it has been pointed out that $\frac{\delta A[\rho]}{\delta_K \rho(x)}$ emerges as the unconstrained derivative of the degree-zero *K*-homogeneous extension of $A[\rho]$ for $\rho_K(x')$ (that is, a $\rho(x')$ of Eq.(2)),

$$\frac{\delta A[\rho]}{\delta_K \rho(x)} = \frac{\delta A[\rho_K^0[\rho]]}{\delta \rho(x)}, \qquad (5)$$

following from two essential conditions, namely, (i) the derivatives of two functionals that are equal over a (K-)restricted domain should also be equal over that domain (K-equality condition), and (ii) for *K*-independent functionals the *K*-constrained derivative should be identical with the unconstrained derivative (K-independence condition). This, for two *K*-constraints, gives

$$\frac{\delta A[\rho]}{\delta_{K_1,K_2} \rho(x)} = \frac{\delta A[\rho_{K_1,K_2}^0[\rho]]}{\delta \rho(x)}, \qquad (6)$$

if $\rho_{K_1,K_2}^0[\rho]$, which is an extension of $\rho_{K_1,K_2}$ that is both $K_1$-homogeneous and $K_2$-homogeneous of degree zero, exists.

For two simultaneous linear constraints, Eq.(4), however, the above idea gives an insufficient basis for the derivation of a constrained differentiation formula, since the extension from $\rho_{L_1,L_2}$ of degree-zero homogeneity (to which degree-zero *L*-homogeneity reduces) is not unique, the extensions $\rho_{L_1,L_2}^0[\rho] = \rho(x) \frac{L_i}{\int g_i(x')\rho(x')dx'}$ (*i*=1,2), e.g., both being homogeneous of degree zero. The problem with the nonunique $\rho_{L_1,L_2}^0[\rho]$ is that it cannot generally yield a $(L_1,L_2)$-constrained derivative formula as $\int \frac{\delta A[\rho_{L_1,L_2}^0[\rho]]}{\delta \rho_{L_1,L_2}^0[\rho](x')} \frac{\delta \rho_{L_1,L_2}^0[\rho](x')}{\delta \rho(x)} dx'$ generally does not fulfill the most substantial requirement, the K-equality condition, for a $\frac{\delta A[\rho]}{\delta_{L_1,L_2}\rho(x)}$. The formulae arising from the two $\rho^0_{L_1,L_2}[\rho]$'s mentioned as examples above even contain only one term instead of two, accounting for the two constraints: $\frac{\delta A[\rho]}{\delta_{L_1,L_2}\rho(x)} = \frac{\delta A[\rho]}{\delta\rho(x)} - \frac{g_i(x)}{L_i}\int \rho(x')\frac{\delta A[\rho]}{\delta\rho(x')}dx'$, which gives just the $L$-constrained derivative formula. The reason for that problem is a kind of degeneracy, namely, the variation of a $\rho^0_{L_1,L_2}[\rho]$ that is $L_1$- and $L_2$-independent (which reduces to a simple $N$-independence) is not necessarily left unrestricted by the $(L_1, L_2)$-constraint, contrary to the case of a single constraint on the functional variable. That can be seen in the case of $\rho^0_{N,L}[\rho] = \rho(x)\frac{N}{\int \rho(x')dx'}$, e.g., in the following way: That on the variation of this $\rho^0_{N,L}[\rho]$, the constraint Eq.(4) alone does not yield any restriction is due to the fact that Eq.(4) allows any $\frac{\rho(x)}{\int \rho(x')dx'}$ but with different ($\int \rho(x)dx$)'s in general, which means that with the addition of the constraint $\int \rho(x)\,dx = N$, the variation of $\rho^0_{N,L}[\rho]$ becomes limited.

In spite of the above degeneracy, and actually, for that very reason, the account for simultaneous linear constraints in functional differentiation can be solved (in general), through following the original way [3,4] to obtain a *K*-conserving differentiation formula, the basis of which was to find an extension $\rho^*_K[\rho]$ (or decomposition, being a matter of approach) that (i) reduces to $\rho_K(x)$ for $\rho_K(x)$ and (ii) fulfills

$$\int f(\rho^*_K[\rho](x))\,dx = K \tag{7}$$

for any $\rho(x)$ (which is a more restrictive condition than degree-zero *K*-homogeneity). Note that those conditions do not yield a unique $\rho^*_K[\rho]$ in the case of linear constraints, since any

$$\rho^*_L[\rho](x) = \rho(x) - \frac{u(x)}{g(x)}\left(\int g(x')\rho(x')dx' - L\right), \tag{8}$$

with a $u(x)$ that integrates to 1, satisfies them (giving the general form (25) in [9], which fulfils only the K-equality condition); however, with the requirement of degree-zero K-homogeneity (that is, simply homogeneity for $K=L$), the ambiguity disappears, $\rho_L^*[\rho]$ becoming $\rho_L^0[\rho]$ (with $u(x) = \frac{g(x)\rho(x)}{\int g(x')\rho(x')dx'}$). For two linear constraints, to find an extension $\rho_{L_1,L_2}^{*0}[\rho]$ that fulfills the 2+1 conditions is a somewhat more difficult task than in the case of a single constraint. It can be solved with the introduction of a function $\sigma(x)$ that integrates to zero:

$$\rho_{L_1,L_2}^{*0}[\rho](x) = \rho(x)\frac{L_1}{\int g_1(x')\rho(x')dx'} - \frac{\sigma(x)/g_1(x)}{\int \frac{g_2(x')}{g_1(x')}\sigma(x')dx'}\left(\frac{L_1}{\int g_1(x'')\rho(x'')dx''}\int g_2(x'')\rho(x'')dx'' - L_2\right).$$

(9)

$\int \frac{g_2(x)}{g_1(x)}\sigma(x)\,dx$ has to be nonzero and $\sigma(x)$ may have a $\rho(x)$-dependence (e.g., some homogeneous $\rho(x)$-dependence) that gives a degree-zero homogeneous $\sigma(x)\left/\int \frac{g_2(x')}{g_1(x')}\sigma(x')dx'\right.$ in $\rho(x)$; otherwise $\sigma(x)$ can be arbitrary. In Eq.(9) the indeces 1 and 2 can be interchanged, yielding an equivalently appropriate extension. Eq.(9), through Eq.(6), leads to the $(L_1, L_2)$-conserving differentiation formula

$$\frac{\delta A[\rho]}{\delta_{L_1,L_2}\rho(x)} = \frac{\delta A[\rho]}{\delta\rho(x)} - \frac{g_1(x)}{L_1}\left\{\int \rho(x')\frac{\delta A[\rho]}{\delta\rho(x')}dx' - \frac{L_2}{\int \frac{g_2}{g_1}\sigma}\int \frac{\sigma(x')}{g_1(x')}\frac{\delta A[\rho]}{\delta\rho(x')}dx'\right\}$$

$$- \frac{g_2(x)}{\int \frac{g_2}{g_1}\sigma}\int \frac{\sigma(x')}{g_1(x')}\frac{\delta A[\rho]}{\delta\rho(x')}dx'. \quad (10)$$

Note that a term $+\frac{\sigma'(x)}{g(x)}\xi\left(\int g(x')\rho(x')dx' - L\right)$, with an arbitrary function $\xi$ for which $\xi(0) = 0$, can also be added to the extension (8); but without any effect on the form (25) in [9] arising from Eq.(8). It may be mentioned that with the choice

$$\sigma(x) = \frac{g_1(x)\rho(x)}{L_1} - \delta(x - x_0) , \qquad (11)$$

Eq.(10) reduces to

$$\frac{\delta A[\rho]}{\delta_{L_1,L_2}\rho(x)} = \frac{\delta A[\rho]}{\delta\rho(x)} - \frac{g_2(x)}{L_2}\int \rho(x')\frac{\delta A[\rho]}{\delta\rho(x')}dx' \qquad (12)$$

if $\frac{g_2(x_0)}{g_1(x_0)} = 0$ and $\frac{1}{g_1(x_0)}\frac{\delta A[\rho]}{\delta\rho(x_0)} = 0$. An essential property of $\frac{\delta A[\rho]}{\delta_{L_1,L_2}\rho(x)}$ given by Eq.(10) is that multiplied by $\rho(x)$ or by $\sigma(x)/g_1(x)$, it integrates to zero.

A more general necessary form for $(L_1, L_2)$-constrained derivatives, coming from the extension

$$\rho^*_{L_1,L_2}[\rho](x) = \rho(x) - \frac{\sigma_1(x)/g_2(x)}{\int \frac{g_1(x')}{g_2(x')}\sigma_1(x')dx'}\left(\int g_1(x')\rho(x')dx' - L_1\right) - \frac{\sigma_2(x)/g_1(x)}{\int \frac{g_2(x')}{g_1(x')}\sigma_2(x')dx'}\left(\int g_2(x')\rho(x')dx' - L_2\right)$$

(13)

(relaxing the homogeneity requirement), is

$$\frac{\delta A[\rho]}{\delta'_{L_1,L_2}\rho(x)} = \frac{\delta A[\rho]}{\delta\rho(x)} - \frac{g_1(x)}{\int \frac{g_1}{g_2}\sigma_1}\int \frac{\sigma_1(x')}{g_2(x')}\frac{\delta A[\rho]}{\delta\rho(x')}dx' - \frac{g_2(x)}{\int \frac{g_2}{g_1}\sigma_2}\int \frac{\sigma_2(x')}{g_1(x')}\frac{\delta A[\rho]}{\delta\rho(x')}dx' , \qquad (14)$$

with $\sigma_1(x)$ and $\sigma_2(x)$ arbitrary functions that integrate to zero and $\int \frac{g_1(x)}{g_2(x)}\sigma_1(x)\,dx \neq 0$ (and $1 \leftrightarrow 2$). The above formula is the most general one that fulfills the most essential condition, namely, the K-equality condition (see above), for a K-constrained derivative, giving back Eq.(10) with

$$\sigma_1(x) = g_2(x)\rho(x) - \frac{\sigma_2(x)g_2(x)/g_1(x)}{\int \frac{g_2(x')}{g_1(x')}\sigma_2(x')dx'} \int g_2(x')\rho(x')dx' . \tag{15}$$

Eq.(13), with Eq.(14), also shows the way for the generalization for an arbitrary number of simultaneous *L*-constraints:

$$\rho^*_{L_1,L_2,\ldots}[\rho](x) = \rho(x) - \sum_i v_i(x)\left(\int g_i(x')\rho(x')dx' - L_i\right) , \tag{16}$$

giving

$$\frac{\delta A[\rho]}{\delta'_{L_1,L_2,\ldots}\rho(x)} = \frac{\delta A[\rho]}{\delta\rho(x)} - \sum_i g_i(x)\int v_i(x')\frac{\delta A[\rho]}{\delta\rho(x')}dx' , \tag{17}$$

with $v_i(x)$'s for which

$$\int g_j(x)v_i(x)\,dx = \delta_{ji} . \tag{18}$$

The construction of such $v_i(x)$'s will be described later, getting help from the derivation of the formulae $\dfrac{\delta A[\rho]}{\delta'_{K_1,K_2,\ldots}\rho(x)}$ with general *K*-constraints.

For two simultaneous general *K*-conservation constraints, of which at least one is nonlinear (more precisely, not homogeneous, as seen later), to derive a constrained differentiation formula, a route based on Eq.(6) will be followed. As $\dfrac{\delta A[\rho]}{\delta_{K_1,K_2}\rho}$ is expected to be equal to the unconstrained derivative of $A[\rho_{K_1,K_2}]$'s degree-zero $K_1$- and $K_2$- homogeneous extension, $A^0_{K_1,K_2}[\rho]$ (*if* that exists),

$$\int \frac{f_i(\rho(x))}{f_i^{(1)}(\rho(x))}\frac{\delta A[\rho]}{\delta_{K_1,K_2}\rho(x)}dx = 0 \tag{19}$$

has to hold for all *i*'s, following from the corresponding relation for $A^0_{K_1,K_2}[\rho]$ (see [9] for details). Eq.(19), utilizing

$$\frac{\delta A[\rho]}{\delta_{K_1,K_2}\rho(x)} = \frac{\delta A[\rho]}{\delta\rho(x)} - f_1^{(1)}(\rho(x))\mu_1 - f_2^{(1)}(\rho(x))\mu_2 \,, \tag{20}$$

then yields two equations for the two multipliers $\mu_i$:

$$\int \frac{f_1(\rho(x))}{f_1^{(1)}(\rho(x))} \frac{\delta A[\rho]}{\delta\rho(x)} dx = K_1\mu_1 + K_{2(1)}\mu_2 \quad \text{and} \quad 1 \leftrightarrow 2 \,, \tag{21}$$

with

$$K_{2(1)} = \int \frac{f_1(\rho(x))}{f_1^{(1)}(\rho(x))} f_2^{(1)}(\rho(x)) dx \quad \text{and} \quad 1 \leftrightarrow 2 \,. \tag{22}$$

The solution of the two equations for $\mu_i$'s yields

$$\mu_1 = \frac{1}{1 - K_{1(2)}K_{2(1)}/(K_1 K_2)} \frac{1}{K_1} \int \left\{ \frac{f_1(\rho(x'))}{f_1^{(1)}(\rho(x'))} - \frac{K_{2(1)}}{K_2} \frac{f_2(\rho(x'))}{f_2^{(1)}(\rho(x'))} \right\} \frac{\delta A[\rho]}{\delta\rho(x')} dx' \quad \text{and} \quad 1 \leftrightarrow 2, \tag{23}$$

which, inserted into Eq.(20), gives the formula for $\frac{\delta}{\delta_{K_1,K_2}}$-derivatives. By construction, $\frac{\delta A[\rho]}{\delta_{K_1,K_2}\rho}$ gives $\frac{\delta A[\rho]}{\delta\rho}$ for functionals $A[\rho]$ independent of $K_1$ and $K_2$. Eq.(23) immediatelly shows why the above formula cannot be applied for two K-constraints with homogeneous $f_i(\rho)$'s (as a consequence of the degeneracy present in that case): for $f_i(\rho)$'s that are $f_i(\lambda\rho) = \lambda^{m_i} f_i(\rho)$, $K_{1(2)} = \frac{m_1}{m_2}K_1$ and $1 \leftrightarrow 2$, which leads to a $\frac{1}{0}$ in the expressions of $\mu_i$'s. (Note that in that case, only one Eq.(19) emerges.)

The extension of the above formula for more than two constraints is straight, on the basis of the above method. However, for that extension, the more general form for $(K_1, K_2)$-constrained derivatives that fulfills only the K-equality condition is worth taking as basis, giving a general frame for all kinds of K-constraints. From the equality of the derivatives

$$\frac{\delta A[\rho]}{\delta'_{K_1,K_2}\rho(x)} = \frac{\delta A[\rho]}{\delta\rho(x)} - f_1^{(1)}(\rho(x))\mu'_1 - f_2^{(1)}(\rho(x))\mu'_2 \tag{24}$$

of two functionals $A_1[\rho]$ and $A_2[\rho]$ at a $\rho(x)$ (which form a *derivative* on a $(K_1, K_2)$-restricted domain has to have; see Sec.2 in [9]),

$$\frac{\delta A_1[\rho]}{\delta \rho(x)} - \frac{\delta A_2[\rho]}{\delta \rho(x)} = f_1^{(1)}(\rho(x))(\mu_1'^{A_1} - \mu_1'^{A_2}) + f_2^{(1)}(\rho(x))(\mu_2'^{A_1} - \mu_2'^{A_2}) \quad (25)$$

follows. Searching for a linear differentiation operator $\dfrac{\delta}{\delta'_{K_1, K_2} \rho}$, Eq.(25), with the introduction of two functions $u_1(x)$ and $u_2(x)$ that integrate to 1 and give $1 - K'_{1(2)} K'_{2(1)} \neq 0$ (see below), leads to

$$\int \frac{u_1(x)}{f_1^{(1)}(\rho(x))} \left( \frac{\delta A_1[\rho]}{\delta \rho(x)} - \frac{\delta A_2[\rho]}{\delta \rho(x)} \right) dx = (\mu_1'^{A_1} - \mu_1'^{A_2}) + K'_{2(1)}(\mu_2'^{A_1} - \mu_2'^{A_2}) \quad \text{and} \quad 1 \leftrightarrow 2, \quad (26)$$

with

$$K'_{2(1)} = \int \frac{u_1(x)}{f_1^{(1)}(\rho(x))} f_2^{(1)}(\rho(x)) \, dx \quad \text{and} \quad 1 \leftrightarrow 2. \quad (27)$$

The solution of the two equations for $(\mu_i'^{A_1} - \mu_i'^{A_2})$'s then yields

$$\mu_1' = \frac{1}{1 - K'_{1(2)} K'_{2(1)}} \int \left\{ \frac{u_1(x')}{f_1^{(1)}(\rho(x'))} - K'_{2(1)} \frac{u_2(x')}{f_2^{(1)}(\rho(x'))} \right\} \frac{\delta A[\rho]}{\delta \rho(x')} dx' \quad \text{and} \quad 1 \leftrightarrow 2, \quad (28)$$

utilizing that the expression for $\mu_i'$ has to be the same for the two functionals. Eq.(24) with $\mu_i'$ determined by Eq.(28) is the generalization of Eq.(25) of [9] for two *K*-constraints, and even for two general constraints $C_i[\rho] = 0$, with the replacement of $f_i^{(1)}(\rho(x))$ with $\dfrac{\delta C_i[\rho]}{\delta \rho(x)}$.

It gives Eqs.(20) and (23) with the choice

$$u_i(x) = \frac{f_i(\rho(x))}{\int f_i(\rho(x')) \, dx'}, \quad (29)$$

and with the transformation from functions $u_i(x)$ integrating to 1 to functions $\sigma_i(x)$ integrating to 0, via

$$u_1(x) = \sigma_1(x)\frac{f_1^{(1)}(\rho(x))}{f_2^{(1)}(\rho(x))} \bigg/ \int \sigma_1(x')\frac{f_1^{(1)}(\rho(x'))}{f_2^{(1)}(\rho(x'))} dx' \quad \text{and} \quad 1 \leftrightarrow 2, \quad (30)$$

it gives the formula that embraces Eq.(14).

Following a similar route as above for three constraints leads to

$$\mu_1' = \frac{1}{1 - K_{1(2)}'K_{2(1)}' - K_{1(3)}'K_{3(1)}' - K_{2(3)}'K_{3(2)}' + K_{1(2)}'K_{3(1)}'K_{2(3)}' + K_{2(1)}'K_{1(3)}'K_{3(2)}'}$$

$$\times \int \left\{ (1 - K_{2(3)}'K_{3(2)}')\frac{u_1(x')}{f_1^{(1)}(\rho(x'))} \right.$$

$$- (K_{2(1)}' - K_{2(3)}'K_{3(1)}')\frac{u_2(x')}{f_2^{(1)}(\rho(x'))} - (K_{3(1)}' - K_{3(2)}'K_{2(1)}')\frac{u_3(x')}{f_3^{(1)}(\rho(x'))} \left. \right\} \frac{\delta A[\rho]}{\delta \rho(x')} dx'$$

$$\text{and} \quad 1 \leftrightarrow i \quad (i=2,3). \quad (31)$$

The essential property

$$\int \frac{u_i(x)}{f_i^{(1)}(\rho(x))} \frac{\delta A[\rho]}{\delta_{K_1,K_2,\ldots}' \rho(x)} dx = 0 \quad (32)$$

holds generally, for arbitrary number of constraints. It can be observed that the multiplier of $\frac{\delta A[\rho]}{\delta \rho(x)}$ in $\mu_i'$ integrates to 1 if multiplied by $f_i^{(1)}(\rho(x))$, and to 0 if multiplied by $f_j^{(1)}(\rho(x))$ ($j \neq i$), giving just $v_i(x)$ in Eq.(16) for linear constraints, hereby this method yielding a general construction of $v_i(x)$'s for an arbitrary number of linear constraints. All $u_i(x)$'s in a formula corresponding to non-homogeneous $f_i(\rho)$'s can be chosen as Eq.(29), fulfilling the K-independence condition (see the second paragraph); while only one of the $u_i(x)$'s corresponding to homogeneous $f_i(\rho)$'s can so be chosen. It is worth underlining here that setting up an extension $\rho_{L_1,\ldots}^{*0}$ of $\rho$ is of relevance even with a derivation of the $(L_1, L_2, \ldots)$-conserving differentiation formula given without the use of it, as the method based on it can be used in complex cases where the straight application of the formula Eq.(17) may not be possible, namely, in the case of functional variables coupled by the constraints present; as in the application of K-constrained differentiation given in [5], where though the problem of the treatment of simultaneous constraints is avoided in the derivation [10] of the constrained derivatives Eqs.(7)-(8). [It is worth mentioning that the constrained derivatives (7) and (8) in [5] can be obtained by the use of Eq.(3) (with $K=L$), that is, of Eq.(17) (with one $v_i(x)$, $v(x) = \rho(x)/L$), as well, by applying it to the single-constraint case of $\dfrac{\delta F_T}{\delta_K \phi_A}$, and taking into consideration that the multipliers of $h(x)$ in Eq.(7) and $\phi_A(x)$ in Eq.(8) in [5], corresponding to the same constraint, must be equal (the multiplier accounting for the other constraint in Eq.(8) is irrelevant with respect to the considered physical theory).]

To treat a time-dependent constraint (2),

$$\int f(\rho(x,t))dx = K(t) , \qquad (33)$$

as well, the above equations need to be modified slightly, replacing $x$ with $x,t$ and excluding $t$ from the integrations. All the formulae derived above, and in [3,4], embrace the discrete case of multi-variable functions, where the functional derivatives in the formulae become partial derivatives with respect to the function variables, and the integrals become summations over the variable indeces, giving e.g.

$$\frac{\partial h(\vec{r})}{\partial_L r_i} = \frac{\partial h(\vec{r})}{\partial r_i} - \frac{l_i}{L} \vec{r} \nabla h(\vec{r}) \qquad (34)$$

for functions of spatial position with constraints

$$\sum_{i=1}^{3} l_i r_i = L . \qquad (35)$$

As another important case of complex constraints, the (single) constraint

$$\prod_{i=1}^{n} \int g_i(x)\rho(x)dx = P \qquad (36)$$

is also worth considering, for which the extension

$$\rho_{L_1 L_2 \ldots}^{*0}[\rho](x) = \rho(x) \sqrt[n]{\frac{P}{\prod_{i=1}^{n} \int g_i(x')\rho(x')dx'}} \qquad (37)$$

leads to the proper ($\prod_i L_i$)-conserving differentiation formula

$$\frac{\delta A[\rho]}{\delta_{L_1 L_2 \ldots} \rho(x)} = \frac{\delta A[\rho]}{\delta \rho(x)} - \frac{1}{n} \sum_{i=1}^{n} \frac{g_i(x)}{L_i} \int \rho(x') \frac{\delta A[\rho]}{\delta \rho(x')} dx' \ . \qquad (38)$$

Finally, it has to be noted that the K-constrained differentiation formulae presented in this paper are valid not only with the unrestricted derivative $\frac{\delta A[\rho]}{\delta \rho(x)}$ but also with the more generally existing corresponding K-restricted derivative $\left.\frac{\delta A[\rho]}{\delta \rho(x)}\right|_{K_1, K_2, \ldots}$ (as detailed in [9] for single constraints), and that the derivatives can be both Fréchet and Gâteaux for linear $K[\rho]$'s, while they are Fréchet in the case of other $K$-constraints.

In summary, the extension of $K$-conserving functional differentiation for an arbitrary number of simultaneous constraints has been presented, completing the method that makes it possible to account for constraints in functional differentiation in a nonvariational way in physical theories.

**Acknowledgements:** The author thanks Nigel Clarke for correspondence. This work was supported by the grant D048675 from OTKA.

[1] P. C. Hohenberg, B. I. Halperin, Rev. Mod. Phys. **49**, 435 (1977);

   L. H. Ryder, *Quantum Field Theory* (Cambridge University Press, 1996).

[2] As, e.g., in N. Clarke, Eur. Phys. J. E **14**, 207 (2004).

[3] T. Gál, Phys. Rev. A **63**, 022506 (2001).

[4] T. Gál, J. Phys. A **35**, 5899 (2002).

[5] N. Clarke, Macromolecules **38**, 6775 (2005).

[6] R. Yerushalmi-Rozen, T. Kerle, J. Klein, Science **285**, 1254 (1999); H. Wang, R. J. Composto, J. Chem. Phys. **113**, 10386 (2000).

[7] R. Balescu, *Equilibrium and Nonequilibrium Statistical Mechanics* (Wiley and Sons, New York, 1975).

[8] A. Oron, S. H. Davis, S. G. Bankoff, Rev. Mod. Phys. **69**, 931 (1997).

[9] T. Gál, J. Math. Chem., doi:10.1007/s10910-006-9216-4 (2007), in press (e-print arXiv:math-ph/0603027).

[10] N. Clarke, private communication (2006).